\documentclass[aps, 
	prd,
	superscriptaddress, 
	preprintnumbers, 
	amsfont,
	amssymb,
	amsmath,
	notitlepage,
	longbibliography,
	nofootinbib]{revtex4-1}
\usepackage[colorlinks=true, 
	linkcolor=blue, 
	citecolor=blue, 
	urlcolor=blue, 
	linktocpage=true]{hyperref}
\usepackage[utf8]{inputenc}
\usepackage[english]{babel}
\usepackage{lmodern}
\usepackage[protrusion=true, expansion=true]{microtype}
\usepackage[dvipsnames]{xcolor}
\usepackage{graphicx}
\usepackage{mathtools}

\definecolor{remFcol}{RGB}{0,140,220}

\begin{document}

\title{Analogue Gravity and Trans-Atlantic Air Travel}
\date{\today} 
\author{Ian G. Moss}
\email{ian.moss@newcastle.ac.uk}
\affiliation{School of Mathematics, Statistics and Physics, Newcastle University, 
Newcastle Upon Tyne, NE1 7RU, UK}

\begin{abstract}
The problem of finding a minimal-time path for an aeroplane travelling in a wind flow
has a simple formulation in terms of analogue gravity. This paper gives an elementary
explanation with equations and some numerical solutions.
\end{abstract}

\maketitle

\section{Introduction}

Anyone flying across the Atlantic cannot fail to notice that the eastbound flight is shorter than the
westbound flight. The more observant will also notice that the flightpath on the eastbound
flight often travels further south than the westbound. The difference in times is due to the
aeroplane taking advantage of the westerly jet stream. 
The inquisitive traveller may ponder what the optimal route may be.
This problem is known as the Zermello navigation problem \cite{Zermello}, and a solution
is built into modern route planning algorithms 
\cite{BijlsmaSakeJ2009OARi,PalopoKee2010WRit,JardinMatthewR2012MfCM}. The aim of the
present paper is to explain the simple formulation of Zermello's problem in terms of ideas from 
analogue gravity \cite{Gibbons:2008zi,Gibbons:2011ib}. (For a review of analogue gravity
see \cite{Barcelo:2005fc}). The discussion is presented in elementary terms that
should be accessible to non-experts in general relativity.

Aeroplanes are designed to fly efficiently at their optimal airspeed, which is typically around
$500\,{\rm kn}$\footnote{Airplane speeds are measured in knots, abbreviated to ${\rm kn}$, equal to
$1.15078\,{\rm mph}$ or $0.51444{\rm m}\,{\rm s}^{-1}$.}. In a wind with velocity ${\bf v}$,
the speed along the ground ${\bf u}$ would have to satisfy the simple relation
\begin{equation}
({\bf u}-{\bf v})^2=c^2,\label{airspeed}
\end{equation}
where $c$ is the airspeed. Later in the paper, this will be shown to be equivalent to finding
paths in curved spacetime with metric
\begin{equation}
ds^2=-(c^2-v^2)dt^2-2v_idx^i dt+g_{ij}dx^idx^j,\label{metric}
\end{equation}
This is the acoustic metric that forms the basis for a description of hydrodynamical waves 
and  is the starting point for analogue models of gravity \cite{PhysRevLett.46.1351}.
The aeroplane follows the same trajectory as the wave-front of a water wave in a flowing
stream.

The object of the exercise is the minimise the time taken to get from point $A$ to point
$B$ at some fixed altitude above the surface of the planet. For this, we appeal to Fermat's 
principle:  the path taken between two points by a ray of light is the path that can be traversed 
in the least time. In the analogue gravity context, this means the optimal route from $A$ to $B$
is defined by the analogue light rays, or null geodesics, in the acoustic metric.

There are other geometric reformulations of Zermello's  problem \cite{Shen2003}.
However, the analogue gravity approach with a good choice of affine parameter along 
the null geodesic helps simplify the equations, and in this respect the analogue gravity approach 
does seem to offer advantages over alternatives, including  the engineering control 
theory methods
\cite{BijlsmaSakeJ2009OARi,PalopoKee2010WRit,JardinMatthewR2012MfCM}.
Specifically, the equations obtained from the analogue gravity approach have a simple
polynomial form.

\section{Methods}

The first simplification is that motion will be restricted to a fixed altitude, and described by 
two coordinates $x^i$, $i=1,2$. 
When explicit coordinates are needed, spherical polar angles $x^i=(\phi,\theta)$ can be used.
Basis vectors will not be normalised, so there is a distinction between
vectors $v^i$ and covectors $v_i$. Tensor indices are lowered
and raised using the metric tensor $g_{ij}$ and its inverse $g^{ij}$. For example, a flight at 
altitude $h$ over a spherical earth of radius $a$ would see local
distances measured by the metric tensor,
\begin{equation}
ds^2=g_{ij}dx^idx^j=r^2(d\theta^2+\sin^2\theta d\phi^2),
\end{equation}
where $r=a+h$.

Analogue gravity arises from introducing an affine parameter $\tau$ along the trajectory, such that
the velocity vector becomes
\begin{equation}
u^i={dx^i\over dt}={dx^i\over d\tau}{d \tau\over dt}.
\end{equation}
The constant airspeed condition can be rewritten as
\begin{equation}
g_{ij}{dx^i\over d\tau}{dx^j\over d\tau}-2v_i{dx^i\over d\tau}{dt\over d\tau}-
(c^2-v^2)\left({dt\over d\tau}\right)^2=0.
\end{equation}
In this form, the condition defines a path $(x^i(\tau),t(\tau))$ in curved spacetime, with zero length in the
acoustic metric (\ref{metric}).
The problem of finding the shortest-time path between two points on the surface of the
Earth is solved by Fermat's principle. A proof of Fermat's principle for a time-independent metric is 
sketched out in Ref. \cite{MTW}, problem {\bf 40.3}. If the metric depends on time, then the null
geodesics either minimise the flight time, or they are extrema of the flight time if the null geodesics
cross one another \cite{0264-9381-7-8-011}.

\begin{figure}
\includegraphics[width=0.5\linewidth]{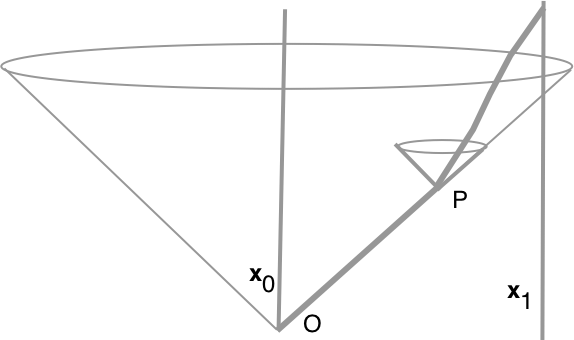}
\caption{Illustration of Fermat's principle. The null geodesics from the event O trace out 
the forward light cone. Any null line from another event P initially lies on the light
cone from P, and cannot pass outside the lightcone from O. 
} \label{fig:fermat}
\end{figure} 

A simple geometrical argument can be used to illustrate the general idea.
Figure \ref{fig:fermat} shows the light rays emanating from an event $O$ at ${\bf x}_0$ and time $t=0$
in a spacetime diagram. These define the forward lightcone for the event $O$. The cone 
intersects the world-line of the destination ${\bf x}_1$ for the first time at time $t$. Any null curve 
from O lies on or inside the forward lightcone of O, and the curve will always intersect the world-line 
of the point ${\bf x}_1$ at a later time than $t$. Where this simple picture breaks down, is ignoring the
possibility that that strong winds can refocus some null geodesics from $O$ to another spacetime 
point, complicating the simple geometry of the lightcone. When this happens, the optimal route is
always one that avoids the refocussing point \cite{0264-9381-7-8-011}.

A standard way to obtain the null geodesics is to use an action principle, starting from the Lagrangian,
\begin{equation}
L=-\frac12(c^2-v^2)\dot t^2-v_i\dot x^i \dot t+\frac12 g_{ij}\dot x^i\dot x^j.
\end{equation}
After introducing the momenta $p_i=\partial L/\partial \dot x^i$ and $p_t=\partial L/\partial (c\dot t)$, the Hamiltonian
constructed form the action is
\begin{equation}
H=-\frac12p_t{}^2-p_t{v^i\over c}p_i+\frac12\left(g^{ij}-{v^iv^j\over c^2}\right)p_ip_j.
\end{equation}
Hamilton's equations are then
\begin{eqnarray}
\dot t&=&-{p_t\over c}-{v^i\over c^2}p_i,\\
\dot x^i&=&g^{ij}p_j-{v^iv^j\over c^2}p_j-p_t{v^i\over c},\\
\dot p_t&=&{p_tp_i\over c^2}{\partial v^i\over\partial t}+{v^j\over c}{p_jp^i\over c^2}{\partial v^i\over\partial t},\\
\dot p_i&=&-\frac12{\partial g^{jk}\over\partial x^i}p_jp_k+{p_t\over c}p_j{\partial v^j\over\partial x^i}
+{p_kv^k\over c^2}p_j{\partial v^j\over\partial x^i}.
\end{eqnarray}
These equations are valid for any analogue spacetime geodesic, but for null geodesics an additional constraint $H=0$ is imposed.
This constraint is the hamiltonian form of the original airspeed condition (\ref{airspeed}). The solutions to the equations 
and the constraint solve the Zermello navigation problem problem of finding the minimum-time paths between the
endpoints  ${\bf x}(0)={\bf x}_0$ and ${\bf x}(\tau)={\bf x}_1$.

A simplified discussion is where the wind velocity is taken to be independent of time.
The time-momentum becomes constant, and up to rescaling of the affine parameter it is possible to set
$p_t=-c$. The remaining equations are as follows,
\begin{eqnarray}
\dot t&=&\gamma,\\
\dot x^i&=&g^{ij}p_j+\gamma v^i,\\
\dot p_i&=&-\frac12{\partial g^{jk}\over\partial x^i}p_jp_k-\gamma p_j{\partial v^j\over\partial x^i},
\end{eqnarray}
where the analogue time dilation factor $\gamma$ between the time and the affine parameter is
\begin{equation}
\gamma=1-{p_iv^i\over c^2}.
\end{equation}
Unlike in relativistic time dilation, the dilation factor can be larger or less than unity.
The constraint $H=0$ becomes, 
\begin{equation}
p^2=\gamma^2 c^2.
\end{equation}
In order to find an optimal path between the endpoints ${\bf x}_0$ and ${\bf x}_1$ numerically, the equations are
integrated from ${\bf x}_0$ with the initial momentum along an arbitrary unit vector $n_i(\alpha)$. 
When substituted into the constraint, this gives a formula for the initial momentum $p_i(0)$,
\begin{equation}
p_i(0)={cn_i(\alpha)\over 1+n_i(\alpha)v^i/c}.
\end{equation}
The distance of closest approach of the geodesic to the final point ${\bf x}_1$ defines a distance function $d(\alpha)$. 
The zeros of the distance function can be found by Newton's method. Each of these zeros represents
a viable null geodesic. The smallest local time at the point of closest approach is the optimal journey time.

Some further simplifying assumptions will be made about the wind velocity field for an illustration of
the general ideas with specific calculations. First of all, the atmosphere will be approximated by an incompressible
gas which is stratisfied into layers at fixed altitude (or more properly air pressure). Such flows
are described as quasi-geostrophic, i.e. Coriolis and pressure gradients are the dominant forces \cite{Cushman}. 
The fluid velocity field is given by a two-dimensional stream function $\psi$, according to
\begin{equation}
v^i=\epsilon^{ij}\partial_j\psi,
\end{equation}
where the $\epsilon^{ij}$ is the alternating tensor, with $\epsilon^{12}=({\rm det}\,g_{ij})^{-1/2}$.
Small oscillations on the stationary flow patterns are known as Rossby waves. Waves with small
wavelengths are advected with the eastward directed flow, but westward movement of the long 
wavelength Rossby waves partially compensates for the underlying drift to leave a relatively
slowly varying wind pattern \cite{Cushman}. 

An exact solution is always useful for testing numerical codes. If the Earth is assumed to be a sphere,
then the stream function $\psi=-v\cos\theta$ gives a soluble set of equations,
\begin{eqnarray}
\dot\theta&=&{p_\theta\over r^2}\\
\dot\phi&=&{p_\phi\over r^2\sin^2\theta}+\gamma {v\over r}
\end{eqnarray}
along with $\dot p_\phi=0$. The constraint closes the system,
\begin{equation}
p_\theta^2+{p_\phi^2\over\sin^2\theta}=\gamma^2 r^2c^2.
\end{equation}
The velocity can be absorbed by introducing a new angular variable $\tilde\phi=\phi-v t/r$, 
and defining a renormalised speed $c'=\gamma c$. The resulting equations describe geodesics
on a sphere, which are the the great circles in the $\theta$ and $\tilde\phi$ coordinates. The
fightpaths are great circles shifted to the east at the speed $v$.

\begin{figure}
\includegraphics[width=0.8\linewidth]{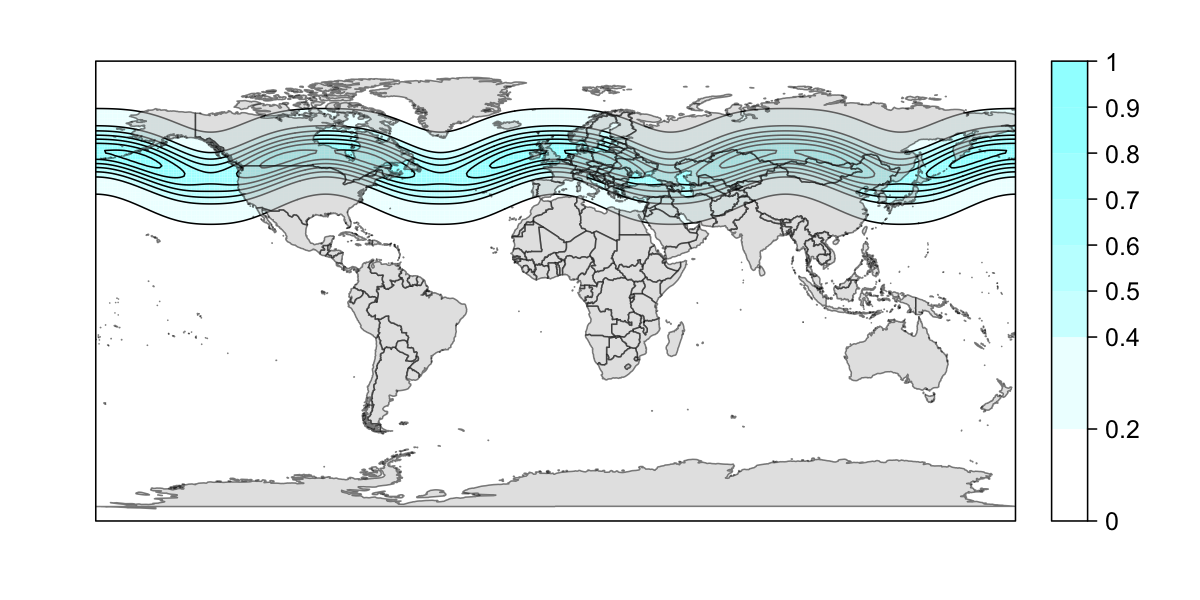}
\caption{Windspeed profile for a simple model of the polar jetstream. The latitude
in Eq. (\ref{stream}) has been set to 50N, the width to 10 degrees, the Rossby wave amplitude to 0.2
and the mode number $m=4$.
} \label{fig:jet}
\end{figure}

For a slightly more realistic model of the jet stream, consider the following stream function
\begin{equation}
\psi=C\arctan\left({\theta-\Theta(\phi)\over w}\right),\quad \Theta(\phi)=\theta_0(1+b\sin\theta\cos (m\phi-\phi_0)).
\label{stream}
\end{equation}
The constant $C$ is fixed by setting the maximum speed of the jet stream. The latitude and width of the stream
are set by $\theta_0$ and $w$, whilst $b$ and $m$ represent the amplitude and the wavelength of the Rosby waves.
An example of the windspeed obtained from this stream function is shown in Fig. \ref{fig:jet}.

\begin{figure}
\includegraphics[width=0.45\linewidth]{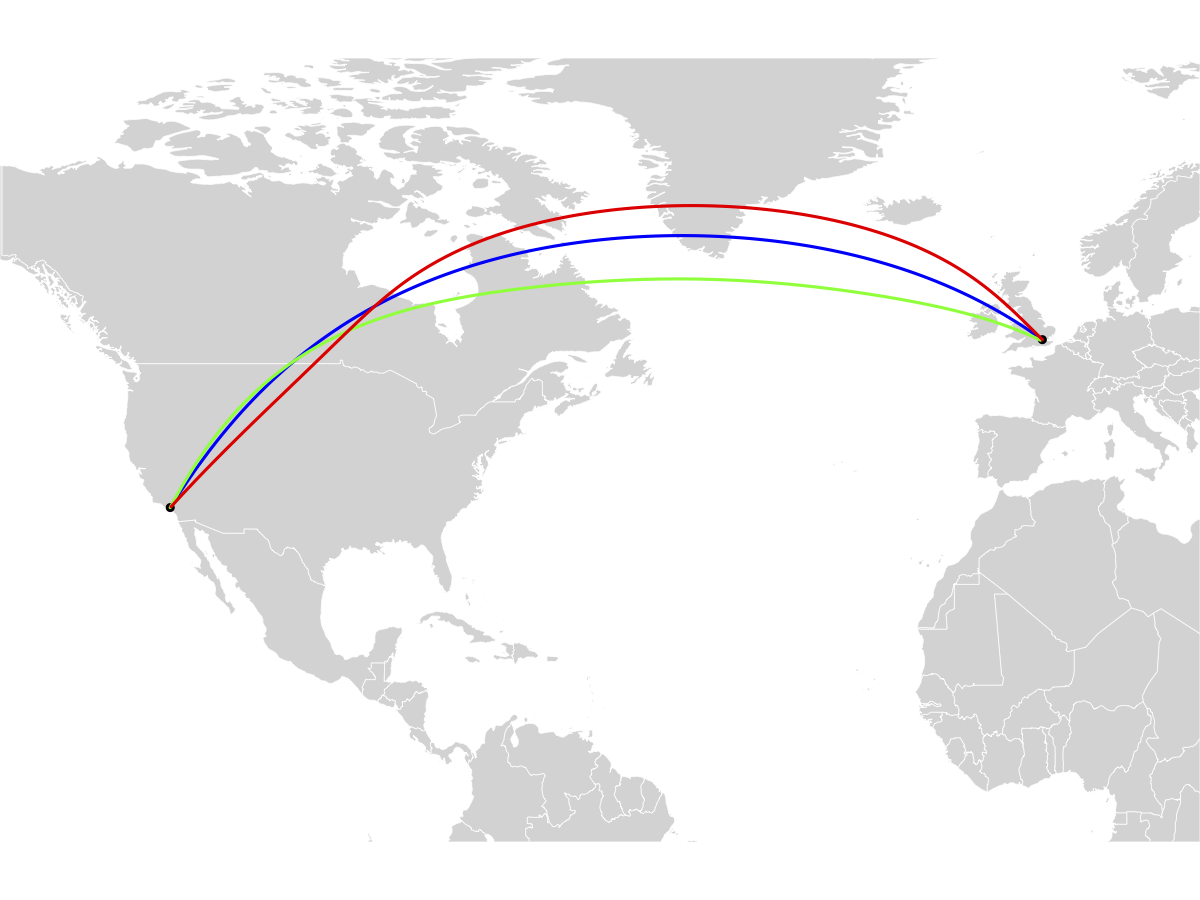}
\includegraphics[width=0.45\linewidth]{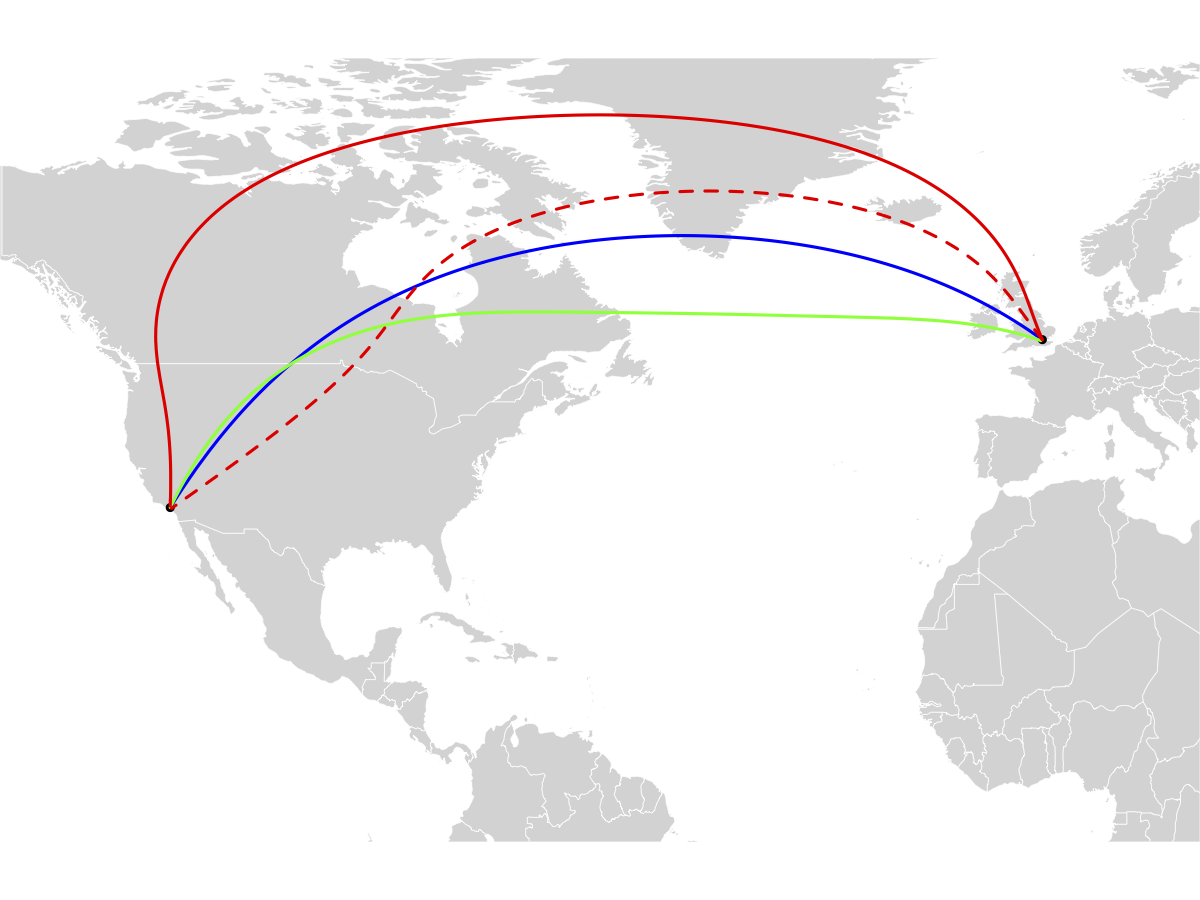}
\caption{Eastbound (green) and westbound (red) optimal paths for a journey between Los Angeles and London
with windspeed $100\,{\rm kn}$ (left) and $180\,{\rm kn}$ (right). The dashed line is a local minimum
of the travel time, the fastest route being the one over the Canadian Rockies. The sensible pilot will 
travel below the powerful jet stream on the westbound route making the westbound paths less relevant.
} \label{fig:paths}
\end{figure}

Figure \ref{fig:paths} shows some of the optimal flightpaths for an aeroplane trajectory in the wind
pattern shown in Fig. \ref{fig:jet} on a spherical Earth. As expected, the eastbound flightpaths follow a more 
southerly route than the geodesic route. The westbound flightpaths are not as useful because then pilot has an option
to fly at lower altitudes where the jet stream is not as strong.  Nevertheless, an interesting phenomenon
arises for wind speed above $100\,{\rm kn}$, when a second route opens up over eastern North America.
This second route is a local minimum of the fight-time.  For strong wind speeds above $140\,{\rm kn}$, the eastern route
becomes the shortest route to LA. 

\begin{figure}
\includegraphics[width=0.45\linewidth]{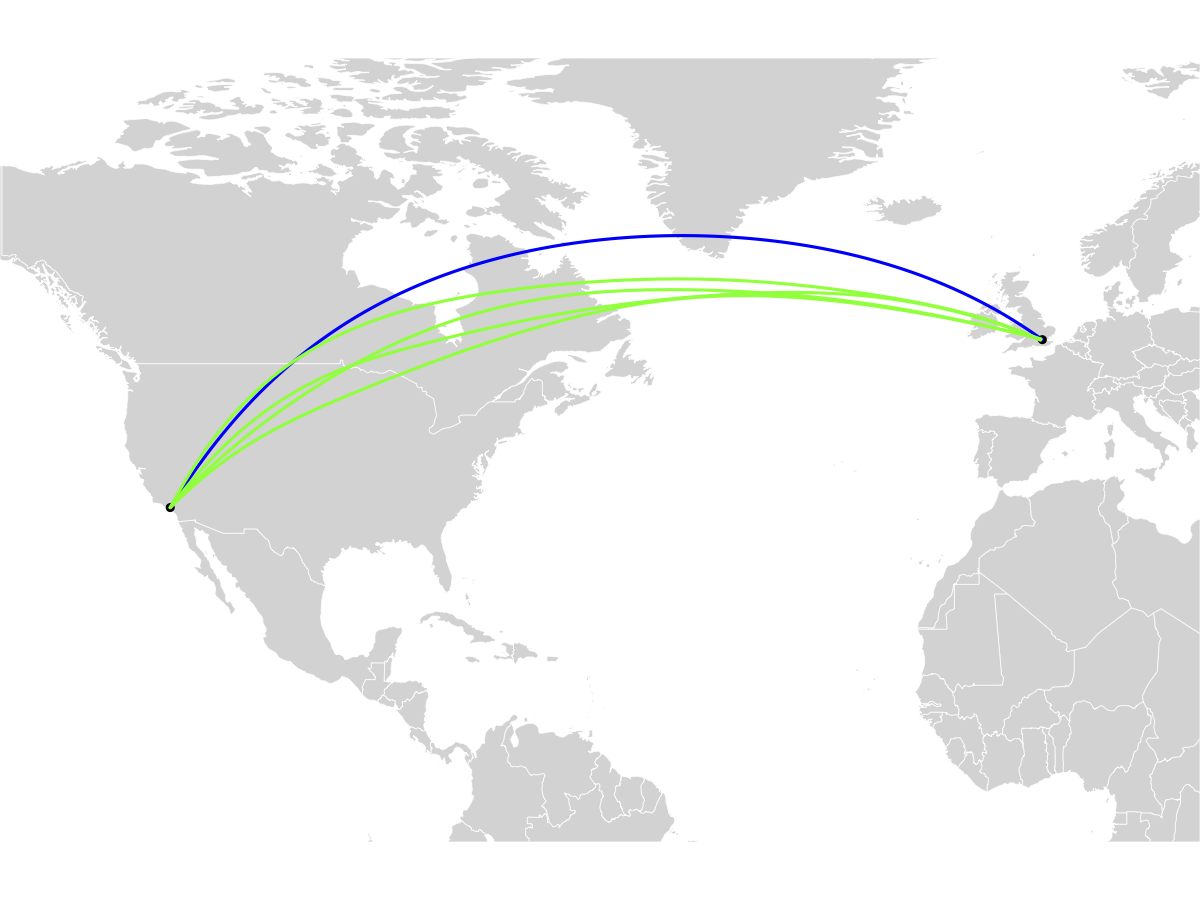}
\includegraphics[width=0.45\linewidth]{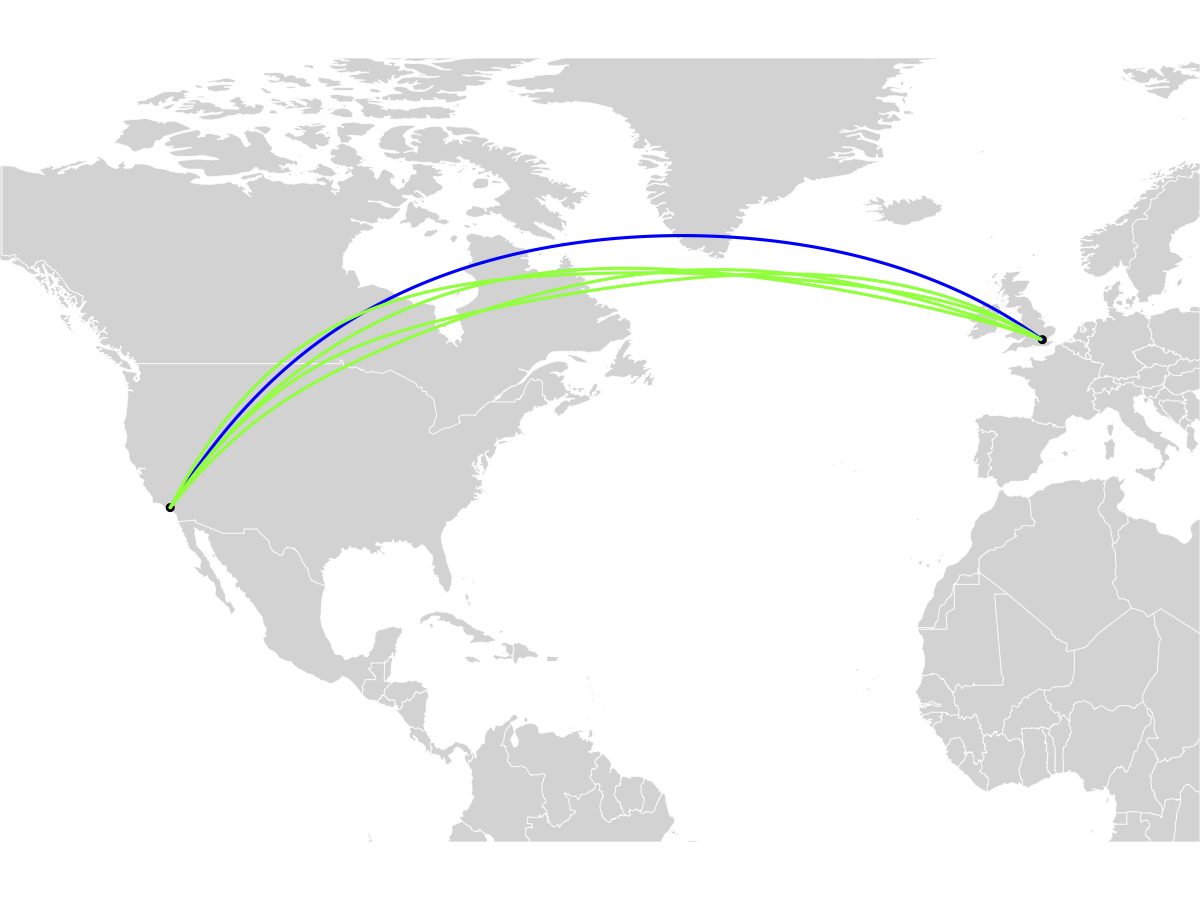}
\caption{Variation of the Eastbound paths for changes in wind patterns with wind speed $100\,{\rm kn}$. 
On the left, the  jet stream latitude $\theta_0$
is 50N, and on the right 55N. Values of the longitude $\phi_0$ differ by 90 degrees. 
} \label{fig:var}
\end{figure}

The detailed flightpaths depend, naturally, on the particular parameters used
to describe the wind pattern. Figure \ref{fig:var} gives an idea of the changes as
a result of the Rossby wave phase $\phi_0$ and the latitude $\theta_0$.

\section{Discussion}

The aim of this paper has been to present the reader with an example of analogue gravity in a `real-world'
setting. Many mathematical idealisations have been used to simplify the analysis and the results are
not meant to be taken seriously as practical solutions to designing aeroplane flight paths. On the other hand,
most of the simplifications used here can easily be improved upon. Replacing the spherical geometry with the
Earth's spheroidal shape is a trivial extension of the results. 
More realistic wind velocity profiles can be included by taking data from existing meteorological sources, either
in grid or spherical harmonic form. Extending the results for time-dependent wind patterns is also
perfectly possible. A harder problem is to take into account the different wind velocity fields is different
strata for the atmosphere. The analogue metric can be extended to apply in three dimensions of space,
but then other factors such as variable airspeed have to be taken into account. Whether this is
would give any improvements on current navigational technology is highly questionable, but hopefully this
account has been informative. 

\acknowledgments
The author is grateful for the encouragement of Gerasimos Rigopoulos to write up this paper, and to
British Airways for an upgrade on the transatlantic flight where the work was begun.
The author is supported by the Leverhulme grant RPG-2016-233, and he acknowledges some support 
from the Science and Facilities Council of the United Kingdom grant number ST/P000371/1.

\bibliography{biblio}

\begin{thebibliography}{12}%
\makeatletter
\providecommand \@ifxundefined [1]{%
 \@ifx{#1\undefined}
}%
\providecommand \@ifnum [1]{%
 \ifnum #1\expandafter \@firstoftwo
 \else \expandafter \@secondoftwo
 \fi
}%
\providecommand \@ifx [1]{%
 \ifx #1\expandafter \@firstoftwo
 \else \expandafter \@secondoftwo
 \fi
}%
\providecommand \natexlab [1]{#1}%
\providecommand \enquote  [1]{``#1''}%
\providecommand \bibnamefont  [1]{#1}%
\providecommand \bibfnamefont [1]{#1}%
\providecommand \citenamefont [1]{#1}%
\providecommand \href@noop [0]{\@secondoftwo}%
\providecommand \href [0]{\begingroup \@sanitize@url \@href}%
\providecommand \@href[1]{\@@startlink{#1}\@@href}%
\providecommand \@@href[1]{\endgroup#1\@@endlink}%
\providecommand \@sanitize@url [0]{\catcode `\\12\catcode `\$12\catcode
  `\&12\catcode `\#12\catcode `\^12\catcode `\_12\catcode `\%12\relax}%
\providecommand \@@startlink[1]{}%
\providecommand \@@endlink[0]{}%
\providecommand \url  [0]{\begingroup\@sanitize@url \@url }%
\providecommand \@url [1]{\endgroup\@href {#1}{\urlprefix }}%
\providecommand \urlprefix  [0]{URL }%
\providecommand \Eprint [0]{\href }%
\providecommand \doibase [0]{http://dx.doi.org/}%
\providecommand \selectlanguage [0]{\@gobble}%
\providecommand \bibinfo  [0]{\@secondoftwo}%
\providecommand \bibfield  [0]{\@secondoftwo}%
\providecommand \translation [1]{[#1]}%
\providecommand \BibitemOpen [0]{}%
\providecommand \bibitemStop [0]{}%
\providecommand \bibitemNoStop [0]{.\EOS\space}%
\providecommand \EOS [0]{\spacefactor3000\relax}%
\providecommand \BibitemShut  [1]{\csname bibitem#1\endcsname}%
\let\auto@bib@innerbib\@empty
\bibitem [{\citenamefont {Zermello}(1931)}]{Zermello}%
  \BibitemOpen
  \bibfield  {author} {\bibinfo {author} {\bibfnamefont {E.}~\bibnamefont
  {Zermello}},\ }\bibfield  {title} {\enquote {\bibinfo {title} {Uber das
  navigationsproblem bei ruhender oder veranderlicher windverteilung},}\
  }\href@noop {} {\bibfield  {journal} {\bibinfo  {journal} {Zeitschrift für
  Angewandte Mathematik und Mechanik}\ }\textbf {\bibinfo {volume} {11}}
  (\bibinfo {year} {1931})}\BibitemShut {NoStop}%
\bibitem [{\citenamefont {Bijlsma}(2009)}]{BijlsmaSakeJ2009OARi}%
  \BibitemOpen
  \bibfield  {author} {\bibinfo {author} {\bibfnamefont {Sake~J}\ \bibnamefont
  {Bijlsma}},\ }\bibfield  {title} {\enquote {\bibinfo {title} {Optimal
  aircraft routing in general wind fields},}\ }\href@noop {} {\bibfield
  {journal} {\bibinfo  {journal} {Journal of Guidance, Control, and Dynamics}\
  }\textbf {\bibinfo {volume} {32}},\ \bibinfo {pages} {1025--1029} (\bibinfo
  {year} {2009})}\BibitemShut {NoStop}%
\bibitem [{\citenamefont {Palopo}\ \emph {et~al.}(2010)\citenamefont {Palopo},
  \citenamefont {Windhorst}, \citenamefont {Suharwardy},\ and\ \citenamefont
  {Lee}}]{PalopoKee2010WRit}%
  \BibitemOpen
  \bibfield  {author} {\bibinfo {author} {\bibfnamefont {Kee}\ \bibnamefont
  {Palopo}}, \bibinfo {author} {\bibfnamefont {Robert~D}\ \bibnamefont
  {Windhorst}}, \bibinfo {author} {\bibfnamefont {Salman}\ \bibnamefont
  {Suharwardy}}, \ and\ \bibinfo {author} {\bibfnamefont {Hak-Tae}\
  \bibnamefont {Lee}},\ }\bibfield  {title} {\enquote {\bibinfo {title}
  {Wind-optimal routing in the national airspace system},}\ }\href@noop {}
  {\bibfield  {journal} {\bibinfo  {journal} {Journal of Aircraft}\ }\textbf
  {\bibinfo {volume} {47}},\ \bibinfo {pages} {1584--1592} (\bibinfo {year}
  {2010})}\BibitemShut {NoStop}%
\bibitem [{\citenamefont {Jardin}\ and\ \citenamefont
  {Bryson}(2012)}]{JardinMatthewR2012MfCM}%
  \BibitemOpen
  \bibfield  {author} {\bibinfo {author} {\bibfnamefont {Matthew~R}\
  \bibnamefont {Jardin}}\ and\ \bibinfo {author} {\bibfnamefont {Arthur~E}\
  \bibnamefont {Bryson}},\ }\bibfield  {title} {\enquote {\bibinfo {title}
  {Methods for computing minimum-time paths in strong winds},}\ }\href@noop {}
  {\bibfield  {journal} {\bibinfo  {journal} {Journal of Guidance, Control, and
  Dynamics}\ }\textbf {\bibinfo {volume} {35}},\ \bibinfo {pages} {165--171}
  (\bibinfo {year} {2012})}\BibitemShut {NoStop}%
\bibitem [{\citenamefont {Gibbons}\ \emph {et~al.}(2009)\citenamefont
  {Gibbons}, \citenamefont {Herdeiro}, \citenamefont {Warnick},\ and\
  \citenamefont {Werner}}]{Gibbons:2008zi}%
  \BibitemOpen
  \bibfield  {author} {\bibinfo {author} {\bibfnamefont {G.~W.}\ \bibnamefont
  {Gibbons}}, \bibinfo {author} {\bibfnamefont {C.~A.~R.}\ \bibnamefont
  {Herdeiro}}, \bibinfo {author} {\bibfnamefont {C.~M.}\ \bibnamefont
  {Warnick}}, \ and\ \bibinfo {author} {\bibfnamefont {M.~C.}\ \bibnamefont
  {Werner}},\ }\bibfield  {title} {\enquote {\bibinfo {title} {{Stationary
  Metrics and Optical Zermelo-Randers-Finsler Geometry}},}\ }\href {\doibase
  10.1103/PhysRevD.79.044022} {\bibfield  {journal} {\bibinfo  {journal} {Phys.
  Rev.}\ }\textbf {\bibinfo {volume} {D79}},\ \bibinfo {pages} {044022}
  (\bibinfo {year} {2009})},\ \Eprint {http://arxiv.org/abs/0811.2877}
  {arXiv:0811.2877 [gr-qc]} \BibitemShut {NoStop}%
\bibitem [{\citenamefont {Gibbons}\ and\ \citenamefont
  {Warnick}(2011)}]{Gibbons:2011ib}%
  \BibitemOpen
  \bibfield  {author} {\bibinfo {author} {\bibfnamefont {G.~W.}\ \bibnamefont
  {Gibbons}}\ and\ \bibinfo {author} {\bibfnamefont {C.~M.}\ \bibnamefont
  {Warnick}},\ }\bibfield  {title} {\enquote {\bibinfo {title} {{The Geometry
  of sound rays in a wind}},}\ }\href {\doibase 10.1080/00107514.2011.563515}
  {\bibfield  {journal} {\bibinfo  {journal} {Contemp. Phys.}\ }\textbf
  {\bibinfo {volume} {52}},\ \bibinfo {pages} {197--209} (\bibinfo {year}
  {2011})},\ \Eprint {http://arxiv.org/abs/1102.2409} {arXiv:1102.2409 [gr-qc]}
  \BibitemShut {NoStop}%
\bibitem [{\citenamefont {Barcelo}\ \emph {et~al.}(2005)\citenamefont
  {Barcelo}, \citenamefont {Liberati},\ and\ \citenamefont
  {Visser}}]{Barcelo:2005fc}%
  \BibitemOpen
  \bibfield  {author} {\bibinfo {author} {\bibfnamefont {Carlos}\ \bibnamefont
  {Barcelo}}, \bibinfo {author} {\bibfnamefont {Stefano}\ \bibnamefont
  {Liberati}}, \ and\ \bibinfo {author} {\bibfnamefont {Matt}\ \bibnamefont
  {Visser}},\ }\bibfield  {title} {\enquote {\bibinfo {title} {{Analogue
  gravity}},}\ }\href {\doibase 10.12942/lrr-2005-12} {\bibfield  {journal}
  {\bibinfo  {journal} {Living Rev. Rel.}\ }\textbf {\bibinfo {volume} {8}},\
  \bibinfo {pages} {12} (\bibinfo {year} {2005})},\ \bibinfo {note} {[Living
  Rev. Rel.14,3(2011)]},\ \Eprint {http://arxiv.org/abs/gr-qc/0505065}
  {arXiv:gr-qc/0505065 [gr-qc]} \BibitemShut {NoStop}%
\bibitem [{\citenamefont {Unruh}(1981)}]{PhysRevLett.46.1351}%
  \BibitemOpen
  \bibfield  {author} {\bibinfo {author} {\bibfnamefont {W.~G.}\ \bibnamefont
  {Unruh}},\ }\bibfield  {title} {\enquote {\bibinfo {title} {Experimental
  black-hole evaporation?}}\ }\href {\doibase 10.1103/PhysRevLett.46.1351}
  {\bibfield  {journal} {\bibinfo  {journal} {Phys. Rev. Lett.}\ }\textbf
  {\bibinfo {volume} {46}},\ \bibinfo {pages} {1351--1353} (\bibinfo {year}
  {1981})}\BibitemShut {NoStop}%
\bibitem [{\citenamefont {Shen}(2003)}]{Shen2003}%
  \BibitemOpen
  \bibfield  {author} {\bibinfo {author} {\bibfnamefont {Z.}~\bibnamefont
  {Shen}},\ }\bibfield  {title} {\enquote {\bibinfo {title} {{Finsler Metrics
  with $K=0$ and $S=0$}},}\ }\href {\doibase 10.4153/CJM-2003-005-6} {\bibfield
   {journal} {\bibinfo  {journal} {Canad. J. Math}\ }\textbf {\bibinfo {volume}
  {55}},\ \bibinfo {pages} {112--132} (\bibinfo {year} {2003})},\ \Eprint
  {http://arxiv.org/abs/math/0109060} {arXiv:math/0109060} \BibitemShut
  {NoStop}%
\bibitem [{\citenamefont {Misner}\ \emph {et~al.}(1973)\citenamefont {Misner},
  \citenamefont {Thorne},\ and\ \citenamefont {Wheeler}}]{MTW}%
  \BibitemOpen
  \bibfield  {author} {\bibinfo {author} {\bibfnamefont {C.W.}\ \bibnamefont
  {Misner}}, \bibinfo {author} {\bibfnamefont {K.S.}\ \bibnamefont {Thorne}}, \
  and\ \bibinfo {author} {\bibfnamefont {J.A.}\ \bibnamefont {Wheeler}},\
  }\href@noop {} {\emph {\bibinfo {title} {Gravitation}}}\ (\bibinfo
  {publisher} {W. H. Freeman, San Fransisco},\ \bibinfo {year}
  {1973})\BibitemShut {NoStop}%
\bibitem [{\citenamefont {Perlick}(1990)}]{0264-9381-7-8-011}%
  \BibitemOpen
  \bibfield  {author} {\bibinfo {author} {\bibfnamefont {V}~\bibnamefont
  {Perlick}},\ }\bibfield  {title} {\enquote {\bibinfo {title} {On fermat's
  principle in general relativity. $i$. the general case},}\ }\href
  {http://stacks.iop.org/0264-9381/7/i=8/a=011} {\bibfield  {journal} {\bibinfo
   {journal} {Classical and Quantum Gravity}\ }\textbf {\bibinfo {volume}
  {7}},\ \bibinfo {pages} {1319} (\bibinfo {year} {1990})}\BibitemShut
  {NoStop}%
\bibitem [{\citenamefont {Cushman-Roisin}\ and\ \citenamefont
  {Beckers}(2011)}]{Cushman}%
  \BibitemOpen
  \bibfield  {author} {\bibinfo {author} {\bibfnamefont {B.}~\bibnamefont
  {Cushman-Roisin}}\ and\ \bibinfo {author} {\bibfnamefont {J.}~\bibnamefont
  {Beckers}},\ }\href@noop {} {\emph {\bibinfo {title} {Introduction to
  Geophysical Fluid Dynamics, Volume 101}}}\ (\bibinfo  {publisher} {Academic
  Press},\ \bibinfo {year} {2011})\BibitemShut {NoStop}%
\end{thebibliography}%

\end{document}